\documentclass[useAMS,usenatbib,usegraphicx]{mn2e}
\usepackage[urlcolor=black,colorlinks=true,linkcolor=black,citecolor=black]{hyperref}
\usepackage{multirow}
\usepackage{color}
\title[IFS of the Homunculus]{Near-infrared integral field spectroscopy of the Homunculus nebula around $\eta$ Carinae using Gemini/{\sc cirpass}\thanks{Based on observations obtained at the Gemini Observatory, which is operated by the Association of Universities for Research in Astronomy, Inc., under a cooperative agreement with the NSF on behalf of the Gemini partnership: the National Science Foundation (United States), the Particle Physics and Astronomy Research Council (United Kingdom), the National Research Council (Canada), CONICYT (Chile), the Australian Research Council (Australia), CNPq (Brazil) and CONICET (Argentina).}}

\author[M. Teodoro et al.] {M.~Teodoro$^{1}$\thanks{e-mail:mairan@astro.iag.usp.br}, A. Damineli$^{1}$, R. G. Sharp$^{2}$, J. H. Groh$^{3}$, C. L. Barbosa$^{4}$\\ $^{1}$Instituto de Astronomia, Geof\'{\i}sica e Ci\^encias Atmosf\'ericas, Universidade de S\~ao Paulo,\\ Rua do Mat\~ao 1226, Cidade Universit\'aria, S\~ao Paulo, 05508-900, Brazil\\ $^{2}$Anglo-Australian Observatory, PO Box 296, Epping, NSW 1710, Australia\\ $^{3}$Max-Planck-Institute f\"ur Radioastronomie, Auf dem H\"ugel 69, D-53121 Bonn, Germany\\ $^{4}$IP\&D, Universidade do Vale do Para\'{\i}ba, Av. Shihima Hifumi 2911, S\~ao Jos\'e dos Campos, 12244-000, SP, Brazil}

\begin{document}

\date{Accepted 2008 March 28. Received 2008 March 20; in original form 2007 May 08}

\pagerange{\pageref{firstpage}--\pageref{lastpage}} \pubyear{2008}

\maketitle

\label{firstpage}

\begin{abstract}

This work presents the first integral field spectroscopy of the Homunculus nebula around $\eta$ Carinae in the near-infrared spectral region ($J$ band). We confirmed the presence of a hole on the polar region of each lobe, as indicated by previous near-IR long-slit spectra and mid-IR images. The holes can be described as a cylinder of height (i.e. the thickness of the lobe) and diameter of $6.5$ and $6.0\times10^{16}$~cm, respectively. We also mapped the blue-shifted component of He\,{\sc i}\,$\lambda$10830 seen towards the NW lobe. Contrary to previous works, we suggested that this blue-shifted component is not related to the \textit{Paddle} but it is indeed in the equatorial disc.

We confirmed the claim of \citet{smith05} and showed that the spatial extent of the \textit{Little Homunculus} matches remarkably well the radio continuum emission at 3~cm, indicating that the \textit{Little Homunculus} can be regarded as a small H\,{\sc ii} region. Therefore, we used the optically-thin 1.3~mm radio flux to derive a lower limit for the number of Lyman-continuum photons of the central source in $\eta$ Car. In the context of a binary system, and assuming that the ionising flux comes entirely from the hot companion star, the lower limit for its spectral type and luminosity class ranges from O5.5\,{\sc iii} to O7\,{\sc i}. Moreover, we showed that the radio peak at 1.7~arcsec NW from the central star is in the same line-of-sight of the `Sr-filament' but they are obviously spatially separated, while the blue-shifted component of He\,\textsc{i}$\lambda$10830 may be related to the radio peak and can be explained by the ultraviolet radiation from the companion star.

\end{abstract}


\section{Introduction}

The Homunculus nebula around the massive luminous blue variable $\eta$~Carinae is the result of an eruptive mass-loss event that took place around 1843, throwing out more than 10~M$_{\odot}$ in a bipolar outflow \citep{morseetal01,gaviola50,ringuelet58,currieetal96,smithetal98a,smithetal98,smithetal03b,smith06}. \citet{hillieretal92} suggested that the lobes of the Homunculus are essentially hollow shells with most of the mass concentrated in two polar caps and thin side walls (see also \citealt{davidsonetal01,smith02,meaburnetal93}). In addition, a ragged equatorial disc also formed, which makes the Homunculus unique among bipolar nebulae \citep{duschletal95}. The Homunculus shows a complex mottled surface with lanes of dust condensation and holes in the lobes \citep{morseetal98}. Although the Homunculus is mainly a reflection nebula, it also has associated intrinsic emission lines due to shocks or photo-excitation \citep{smith05,smith04,smith06,smith02,davidsonetal01,allenetal91,allenetal93,hillieretal92}. Such studies showed that the observed velocity pattern of a spectral line along the Homunculus can be used to determine the origin of each component of the emission.

Surprisingly, another bipolar structure lying inside the Homunculus was discovered by \citet{ishibashietal03} using long-slit spectroscopy obtained with the Space Telescope Imaging Spectrograph (STIS) aboard the Hubble Space Telescope (\textit{HST}). The so-called \textit{Little Homunculus} was detected in more than 30 optical lines of [Fe\,{\sc ii}] and [Ni\,{\sc ii}] (such as [Fe\,{\sc ii}]~$\lambda$4891, $\lambda$4907, $\lambda$4975, and [Ni\,{\sc ii}]~$\lambda$7380) as well as in He\,{\sc i} and H\,{\sc i} lines \citep{ishibashietal03}, and is particularly bright in the near-IR  [Fe\,{\sc ii}] lines \citep{smith02}. The proper motion and radial velocities analysis are consistent with the \textit{Little Homunculus} forming in a smaller mass-ejection event that occurred around 1890 \citep{ishibashietal03,smith05}. \citet{smith05} estimated an ejected mass and kinetic energy for the \textit{Little Homunculus} of 0.1--0.2~M$_{\odot}$ and roughly 10$^{47}$~erg, respectively -- values which are at least two orders of magnitude lower than those for the Great Eruption that created the larger Homunculus \citep{smithetal03b}. Although \citet{smith05} mapped the basic structure of the \textit{Little Homunculus} using five slits oriented parallel to the major axis and separated by $\approx$1 arcsec, the spatial distribution of the \textit{Little Homunculus} with a higher angular resolution \textit{in the near-infrared} has not been done so far.

$\eta$~Car is also surrounded by a broken toroidal ring structure \citep{smithetal00,smithetal02,morrisetal99,ishibashietal03}, which absorbs part of the ultraviolet radiation, while allowing the rest to escape through holes and excite/ionise the surrounding gas at large distances from the central source \citep{smith06}. An excellent example of this effect is given by the blue-shifted component of the He\,{\sc i}~$\lambda$10830 line projected onto the NW lobe \citep{smith02}.

This work presents the results of the first integral-field spectroscopy mapping of the Homunculus nebula. The paper is organized as follows. The observations and data reduction are described in \S\ref{obs}. A discussion about the line-formation mechanism throughout the nebula around $\eta$~Carinae is presented in \S\ref{mec}, while the results are shown in \S\ref{res}. In \S\ref{dis} is a comparison of the continuum radio-emission and our results as well as our estimates about the properties of the hot-companion star. Finally, our conclusions are summarized in \S\ref{sum}.

\section{Observations and data reduction}\label{obs}

The integral field observations of $\eta$~Car were recorded on 2003 March 14, 15 and 18 at the 8m Gemini South telescope using the visitor instrument CIRPASS\footnote{\href{http://www.ast.cam.ac.uk/~optics/cirpass/cirpass\_index.html}{www.ast.cam.ac.uk/$\sim$optics/cirpass/cirpass\_index.html}}, a spectrograph developed by the Cambridge instrumentation team \citep{parryetal04}. CIRPASS has 490 hexagonal lens placed at the integral field unit with a spatial sampling interval of 0.25 arcsec per lens. It provides a wavelength coverage of $\lambda\lambda$10620--12960, with a resolving power ($\frac{\lambda}{\Delta\lambda}$) of 3200.

The observations were originally planned for two epochs, one at the high- (2003 March) and other at the low-excitation state (2003 July) of the 5.5 yr cycle \citep{damineli96, daminelietal00}. Unfortunately, the observing run during the low excitation state was lost due to poor weather conditions.

The following strategy was adopted to map the whole Homunculus: 2 images were taken at a given position of the nebula, and then the IFU was shifted by 0.88 arcsec along the North--South axis to take the next 2 images. The final dataset comprised 2 images from 44 IFU positions, which corresponds to a final mosaic of 6299 spectra and covers the whole Homunculus.

The data were reduced using standard near-infrared techniques. First, any spurious features were removed by the {\sc iraf}\footnote{{\sc iraf} -- Image Reduction and Analysis Facility -- is written and supported by the {\sc iraf} programming group at the National Optical Astronomy Observatories (NOAO) in Tucson, Arizona. NOAO is operated by the Association of Universities for Research in Astronomy (AURA), Inc. under cooperative agreement with the National Science Foundation}/{\sc cosmicrays} task. After that, the spectra were flat-fielded, and an optimal spectral extraction was performed in order to account for the overlapping wings among neighboring spectra. The wavelength calibration was done using an argon lamp spectra and a polynomial interpolation (RMS $\sim 0.1$~{\AA}). The telluric lines were removed dividing the Homunculus spectra by the spectrum of a hot early-type star observed just after the science observation. The photospheric lines of the telluric standard was removed previously through an interpolation of the adjacent continuum. The FWHM of the point-spread function of our data is 0.4~arcsec, which was measured using the intensity profile of the standard star. Finally, a full data cube was constructed from the individual spectra using our own tasks written in {\sc idl}. The signal-to-noise ratio (S/N) is quite variable along the entire mosaic: within the region where the Homunculus nebula is located, the average S/N in the continuum is $\approx 25$, while throughout the outer ejecta it is roughly 10.

\section{Mechanisms of excitation and/or ionization in the circumstellar ejecta of $\eta$~C\lowercase{ar}}\label{mec}
In this paper, we focus on two emission lines found throughout the Homunculus, namely, [Fe\,{\sc ii}]~$\lambda$12567 and He\,{\sc i}~$\lambda$10830. As usual in the spectra of the Homunculus, these lines show many components of intrinsic and reflected emission, which can be used to map the spatial distribution of structures lying inside or outside the Homunculus. Hence, it is very important to know about the mechanisms of line-formation in order to better understand the physical conditions of the emitting and reflecting regions.

In this section we will discuss about the process of line-formation inside and outside the Homunculus nebula. We must stress that throughout this paper, we refer to `photo-excitation' as the process by which a photon of the radiation field is absorbed by an atom or ion of the gas with the promotion of one electron to a higher energy level without ionization (bound-bound transition). If the incident photon has an energy greater than the ionization potential of the atom or ion, then one electron will be stripped out of it. This process is called `photo-ionization' and is responsible for bound-free transitions.

On the other hand, if the gas is hot enough to keep most of the atoms ionized, as usually found around massive stars, then the electronic density will be high and probably the electrons will collide with the ions and excite their electron to an upper level (more energetic). This process is called `collisional-excitation' or `collisional-ionization', depending on the capability of the incident electron in removing or not one electron from the target ion.

The return of the electron to the ground state of the ion -- known as recombination -- will be followed by the emission of many photons with different energy, which gives rise to the emission-line spectrum of some stars.

\subsection{The Homunculus nebula}
Many studies have shown that in the lobes of the Homunculus nebula, photo-excitation is the predominant mechanism for populating the atomic levels of the ions \citep{gulletal05,verneretal05,smith06}. This is because the strong stellar radiation is absorbed by dust in the nebula which, in turn, is heated to a few hundreds degrees Kelvin and starts to emit a reprocessed radiation field responsible for keeping the observed ionization structure inside the walls \citep{smithetal07ii}. However, this is not ionizing radiation since the central source has a dense stellar wind which absorbs nearly all of the Lyman-continuum photons. The few photons that escape the stellar wind are absorbed either by the toroidal structure at the equator or by the \textit{Little Homunculus}, which causes strong variability in the radio continuum \citep{duncanetal97}.

On the other hand, the stellar wind shows a latitude-dependent profile, being more dense and fast in the polar regions than in the equatorial region \citep{smithetal03}. The expansion velocity of the Homunculus is about 600~km\,s$^{-1}$ while the stellar wind reaches terminal velocities in the range of 600 to 1000~km\,s$^{-1}$. Thus, inside the lobes it is expected to detect collisionally-excited emission lines as well. Indeed, the strength of IR [Fe\,\textsc{ii}] and H$_{2}$ lines in the NW lobe relative to SE suggest that there is a combination of slow shocks and photo-excitation \textit{inside the lobes}. The shocks are needed to explain the observed value $\ga$~35 for the ratio of [Fe\,\textsc{ii}]~$\lambda$16435 to Br$\gamma$ \citep{smithetal01}, though it should not exceed the threshold velocity either for dissociation of H$_{2}$ -- detected by \citealt{smithetal01} -- or to emit any hard X-ray photon -- as noted by \citealt{weisetal01}. Therefore, the walls of the nebula must be composed of dense, small neutral/molecular knots or clumps \citep{morseetal98} so that the fast bi-polar wind could escape into the outer ejecta without strong interaction with the Homunculus (similar to the well-known Rayleigh-Taylor instabilities).

Therefore, in the lobes of the Homunculus nebula, there is a competition between photo-excitation and collisions as the main line-formation mechanism, the former being the most significant excitation/ionisation process.

\subsection{The equatorial region}
In the equatorial region, however, the stellar wind is slower than in the polar regions. It also presents a lower density and consequently, the emission due to collisions is weak. Therefore, the equatorial region is largely dominated by photo-excitation. This is because of both the proximity to the central source and the high ionization flux from the equatorial region of the central star \citep{smithetal03}. The presence of a torus around the system is revealed by narrow-band IR images and confirmed at radio wavelengths \citep{smithetal98,duncanetal97} as well as in emission lines. Nevertheless, this structure is not continuous but shows either dense clumps (where low-ionization ions are detected) and holes (through where radiation can escape). Examples of these structures are the so-called `Sr-filament' \citep{zethsonetal99,gulletal00,gulletal01,zethsonetal01,bautistaetal02,hartmanetal04,bautistaetal06} and the He\,\textsc{i}~$\lambda$10830 emission columns \citep{smithetal02}.

\subsection{The outer ejecta}
The picture drastically changes when considering the line-formation process outside the Homunculus nebula, namely, in the outer ejecta. This region is nitrogen-rich \citep{smithetal04a} and responsible for practically all of the observed X-ray flux up to 1.5~keV, which implies shock velocities in excess of 1500~km\,s$^{-1}$\citep{weisetal04}. These shocks are sufficient to excite (and even ionize) ions to higher energy levels than those observed in the Homunculus. This is supported by the observation of N\,\textsc{vi}/\textsc{vii}, Si\,\textsc{xiii}/\textsc{xiv}, Mg\,\textsc{xii} in the X-ray part of a thermal-emission spectrum and strong N\,\textsc{ii}, [O\textsc{iii}], [S\,\textsc{iii}] and Si\,\textsc{ii} and many other lines of high-ionization energy ions in the optical range \citep{smithetal04,weisetal04,hamaguchietal07}.

Thus, throughout the Homunculus nebula the process of line-formation is mainly via photo-excitation with a little contribution from slow shocks inside the lobes at high stellar latitudes, while in the outer ejecta the predominant process is excitation/ionisation via collisions, with little contribution from photo-excitation.

\section{Results: $J$-band spatial maps of the nebula around $\eta$~C\lowercase{ar}}\label{res}

\subsection{Structures in the [Fe\,{\sc ii}]~$\lambda$12567 line}\label{collisions}

This section presents the spatial structure and kinematics of the photo-excited regions found in the Homunculus. For the first time, a complete spatial map of such photo-excited regions is presented in the near-infrared. Although velocity maps were already made using forbidden lines in the optical with higher spatial resolution using \textit{HST} \citep{ishibashietal03}, the use of the near-infrared region in this work allows the observer to peer through the circumstellar dust and probe the environment around $\eta$~Car.

Extensive work has been done in the near-infrared by N. Smith \citep{smith02,smith04,smith05,smith06} to establish the overall kinematics of the regions around $\eta$~Car, in special the \textit{Little Homunculus} \citep{smith05}. However, a complete spatial map has not been done so far in near-infrared. The IFU observations presented here should therefore be interpreted as complementary information to what has been presented by previous works using long-slit spectroscopy.

\begin{figure}\centering \includegraphics[width=8.4cm]{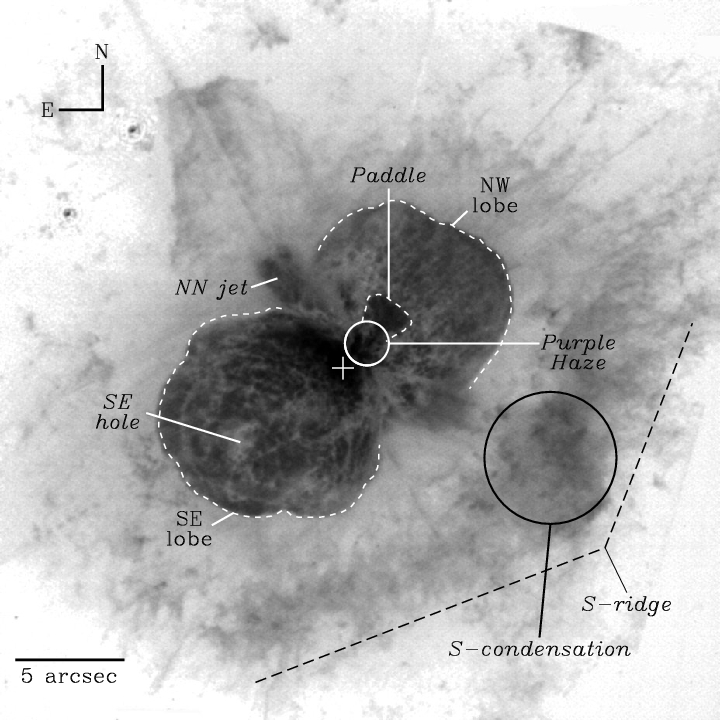} \caption{\label{fig1}Optical \textit{HST}/WFPC2 image of the ejecta around $\eta$~Car (bottom image of fig. 3 from \citealt{morseetal98}, reproduced by permission of the AAS). This image shows the adopted nomenclature of some of the most important regions around $\eta$~Car.} \end{figure}

Fig.~\ref{fig1} is an image of the circumstellar environment of $\eta$~Car obtained with the Planetary Camera of the Wide Field Planetary Camera 2 (WFPC2) on board the \textit{HST}. It is a negative of the image shown in fig. 3 of \citet{morseetal98}, reproduced by permission of the AAS. This image is referred to throughout this paper for the location and standard nomenclature of specific regions in the ejecta of $\eta$~Car.

\begin{figure}\centering \includegraphics[width=8.4cm]{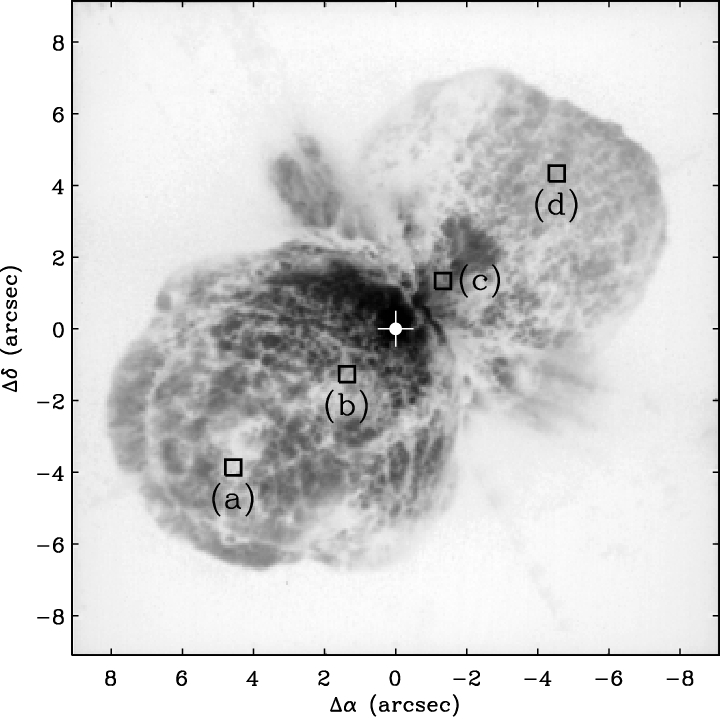}
\includegraphics[width=8.4cm]{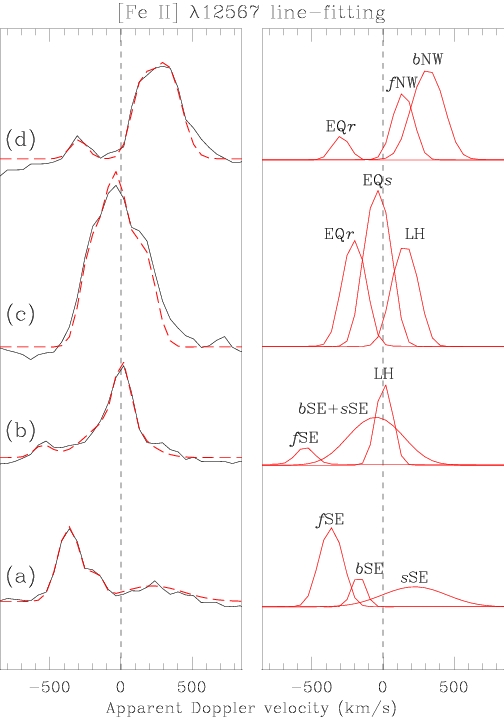} \caption{\label{fig2}The upper panel (background image from fig. 2 of \citealt{morseetal98}) shows the position where we extracted the spectra shown in the lower panel. Although the boxes in the upper panel have approximately the dimension of one lens (0.25~arcsec), each spectrum in the lower panel is a median-combined of 4 adjacent lens in order to improve the signal to noise ratio. The abbreviations is described in \S\ref{idspec}.} \end{figure}

\subsubsection{Identifying line components in the Homunculus spectrum}\label{idspec}

We used the kinematic model of the Homunculus proposed by \citet{davidsonetal01} to identify reflected and intrinsic emission components. This model was built using long slit observations of forbidden spectral lines in the optical, mapping the emission coming from inside the Homunculus with spatial and spectral resolution of 0.1~arcsec and $\approx90$~km\,s$^{-1}$, respectively. Thus, as a first approximation, the \citet{davidsonetal01} model is accurate enough (for our objectives) to classify the observed components of a given line as intrinsic or reflected.

Typical spectra from the Homunculus are presented in Fig.~\ref{fig2}. The adopted abbreviations are as follows: $f$ ($b$) means intrinsic emission coming from the {\it front} ({\it back}) wall of the SE or NW lobe and $s$ means {\it scattered} emission reflected in the SE or NW lobe. As usual, we refer to $front$ ($back$) wall as the near (far) side of each lobe of the Homunculus. Emission associated with the equatorial ejecta is labelled as either EQ$s$ or EQ$r$, where $s$, in this case, means {\it slow}- and $r$, {\it rapid}-moving gas.

We identified all of the [Fe\,\textsc{ii}]~$\lambda$12567 components in order to map the emission structures of the Homunculus. The Doppler velocity of each component mentioned hereafter is heliocentric and was obtained by a multi-gaussian line fitting procedure. The components are as follows:

\begin{itemize}
\item $f$SE ($\approx-400$~km~s$^{-1}$) is an intrinsic emission due to Balmer excitation to an upper level followed by recombination to a lower level (bound-bound transition) or recombination of Fe$^{++}$ to Fe$^{+}$ (free-bound transition) inside the SE lobe;
\item $b$SE ($\approx-200$~km~s$^{-1}$) is intrinsic emission coming from the back wall of the SE lobe and is due to recombination of Fe$^{++}$ to Fe$^{+}$ as well;
\item $s$SE ($\approx+200$~km~s$^{-1}$) is emission from the stellar wind scattered in the front wall of the SE lobe;
\item EQ$s$ ($\approx-100$~km\,s$^{-1}$) is due to intrinsic emission from slow-moving material located in the equatorial plane, which intercepts our line-of-sight towards the NW lobe. This component was also detected in [Ni~II]~$\lambda$7380 by \citet{davidsonetal01} and in many low-ionization emission lines such as [Mn\,\textsc{ii}], [Cr\,\textsc{ii}] and [Ti\,\textsc{ii}]. It is due to the `Sr-filament' discussed in \S~\ref{radio};
\item EQ$r$ ($\approx-250$~km\,s$^{-1}$) is due to equatorial ejecta illuminated directly by ionizing radiation from the central source. This component comes from a region known by its strong continuum radio emission and variability. We will discuss this feature in \S~\ref{radio};
\item $f$NW ($\approx+100$~km~s$^{-1}$) is an intrinsic emission associated with the near wall of the NW lobe. It is due to the same mechanism as $f$SE;
\item $b$NW ($\approx+300$~km~s$^{-1}$) is also intrinsic emission related to recombination at the polar region of the NW lobe;
\item LH is emission associated with the \textit{Little Homunculus}.
\end{itemize}

\subsubsection{The hole at the pole of lobes}\label{hole}

\begin{figure*}\centering \vbox {\vfil \includegraphics[width=16cm]{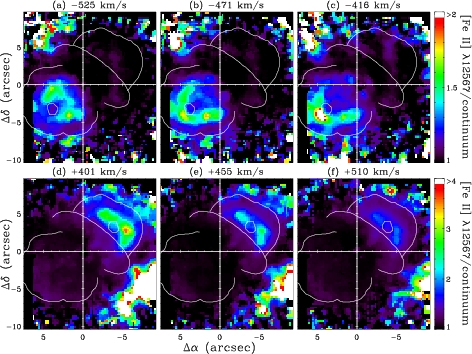} \caption{\label{fig3}Velocity maps of the [Fe\,{\sc ii}]~$\lambda$12567 line showing a hole both in the SE and NW lobe \citep{smithetal98}. The coordinates of the NW hole are the same as the SE hole but rotated by 180\degr around the centre of the axis. Notice the change in the color scale to bring up weaker emissions in each lobe. The emission from the surrounding of the hole is almost 50 per cent greater than that coming from the hole itself. The system velocity is 8.1 km\,s$^{-1}$ heliocentric \citep{smith06}. A full animated version of this figure is available at \href{http://www.astro.iag.usp.br/~damineli/feii_holes.gif}{www.astro.iag.usp.br/$\sim$damineli/feii\_holes.gif} with the same color scale.}\vfil} \end{figure*}

We detected two regions where emission due to recombinant Fe$^{++}$ is nearly absent. These regions can be seen in Fig.~\ref{fig3}, which shows velocity maps of [Fe\,{\sc ii}]~$\lambda$12567 from $-525$~km~s$^{-1}$ to $+510$~km~s$^{-1}$. On both lobes, a small polar region with extremely reduced emission in this line can be identified (circular contour in each lobe in Fig.~\ref{fig3}). The flux from these holes are typically $\sim50$ per cent lower than its immediate surroundings. Moreover, such deficit of [Fe\,{\sc ii}] and H$_2$ emission at these locations was also reported by \citet{smithetal98} and \citet{smith06}. 

It is tempting to associate the hole seen in the [Fe\,{\sc ii}]~$\lambda$12567 velocity maps with the structure known as \textit{SE hole} (see Fig.~\ref{fig1}), observed in optical and near-infrared wavelengths \citep{smithetal98}. To investigate that hypothesis, we compared the relative position between the \textit{SE hole} and the SE region with low emission of [Fe\,{\sc ii}]~$\lambda$12567. Besides, we also compared the position of those regions with the location of the lobe's pole\footnote{By the word `pole' we mean to say the location in the lobe where the stellar latitude is $90\degr$. Obviously, this position is model-dependent.}. To do so, we assumed the kinematic model of the Homunculus of \citet{smith06}, which was obtained by tracing the H$_2$ emission along the nebula and gives a more accurate position for the location of the pole.

Our results give strong support to the idea that there indeed is a hole at the pole of each lobe. The facts that led us to this conclusion are mainly two:

\begin{itemize}
\item the location of the SE region with lower emission of [Fe\,{\sc ii}]~$\lambda$12567 does not match neither the position of the \textit{SE hole} nor the pole of the lobe. Instead, they are shifted from one another by $\approx 0.7$~arcsec along the major axis of the Homunculus, i.e. the \textit{SE hole} is half-way between the pole of the lobe and the region with weak emission of [Fe\,\textsc{ii}]~$\lambda$12567. Hence, the lack of Fe$^{+}$ emission is not associated with either the \textit{SE hole} or the pole.

\item the absence of thermal infrared emission from dust reported by \citet{smithetal98} is also a strong indication that those polar holes are indeed lower-density regions, and not shadows.
\end{itemize}

[Fe\,\textsc{ii}]$\lambda$12567 is most likely to arise from a warm, low-density region \textit{inside} the lobe because when the stellar radiation field penetrates the wall of the lobes -- which has a hydrogen density of about $10^{7}$~cm$^{-3}$ -- it gets more attenuated and then Fe$^{+}$ recombines to Fe$^{0}$ and we see no more emission from Fe\,\textsc{ii} \citep{smithetal07ii}. Thus, the spatial distribution of this line represents the emission coming from the inner part of the lobes. Together with the fact that molecular hydrogen emission comes from a region of the lobe that is shielded from strong radiation -- i.e. just outside it --, we could get a rough estimate of the lobes' thickness.

A first approximation of the geometry of the hole is to consider it as a cylinder with linear diameter\footnote{To conversion between apparent and linear size, we adopted a distance of 2.25~kpc to $\eta$~Car \citep{davidsonetal01}.} \textit{d} -- the diameter of the [Fe\,{\sc ii}] non-emitting region -- and height $\Delta$R$_0$ -- the distance between the SE pole of the Homunculus and the [Fe\,{\sc ii}] non-emitting region. Hence, considering an inclination angle of \textit{i}=41\degr from the line-of-sight \citep{davidsonetal01,smith06}, we obtained for the height and diameter of the cylinder, respectively, a linear size of $6.5\pm0.4\times10^{16}$ and $6.0\pm0.3\times10^{16}$~cm. The errors quoted here are due only to our uncertainty in position and do not include the uncertainty in distance, which is in the range of 0.1--$2\times10^{16}$~cm for most studies \citep{davidsonetal01,smith06}.

\begin{figure*}\centering \vbox {\vfil \includegraphics[width=16.8cm]{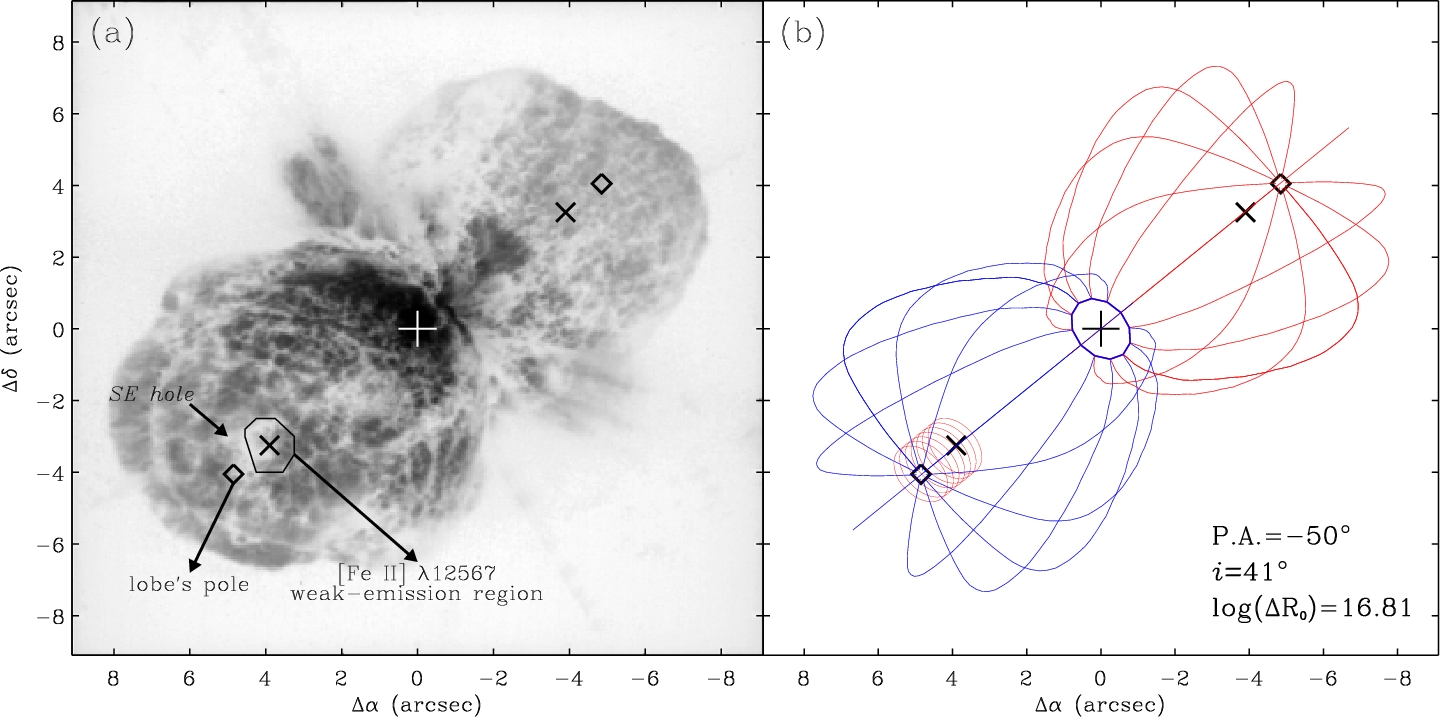} \caption{\label{fig4}(a) Position of the features used to estimating the thickness of the lobes: $\diamond$ marks the location of the pole of the lobe (as in \citealt{smith06}), while $\times$ indicates the centre of the [Fe\textsc{ii}]~$\lambda$12567 weak-emission region (circular contour). The greyscale background image is from \citet{morseetal98}. (b) A cylinder in the SE lobe superimposed on the Homunculus model. From the $\times$ mark to the $\diamond$ symbol, the linear height of the cylinder is roughly $6.5\times10^{16}$~cm (log($\Delta\rm{R_0}$)=16.81), which we assumed as the thickness of the polar region of the lobes. The diameter of the cylinder is assumed to be the same as that observed for the [Fe\textsc{ii}]~$\lambda$12567 weak-emission region, which has a linear size of $6.0\times10^{16}$~cm. The adopted values for the inclination and position angle of the Homunculus model are also given (the same parameters are applied to the cylinder as well).}\vfil} \end{figure*}

The right panel of Fig.~\ref{fig4} shows the adopted model for the height and radius of the hole in the lobes of the Homunculus. The coordinates of the non-emitting region in the NW lobe were obtained by considering the position of the same region in the SE lobe, mirrored relatively to the central star by 180\degr.

We also estimated the thickness of the lobes at lower latitudes by measuring the `limb-darkening' profile seen in our velocity maps (cf. Fig.~\ref{fig3}). Indeed, there is a clear separation between the [Fe\,\textsc{ii}]$\lambda$12567 emission and the optical limit of the Homunculus' lobes because of the high-density medium inside the wall of the lobes. The observed mean separation in both lobes is 1.3~arcsec, which corresponds to a linear thickness of $4.4\pm0.5\times10^{16}$~cm (here, the errors are due to the irregularity of the [Fe\,\textsc{ii}]$\lambda$12567 emission region). This result suggests that the thickness of the lobes also presents a latitude-dependent effect, which makes it almost 50 per cent thicker at polar regions than at lower latitudes.

Both holes -- SE and NW -- define an axis with position angle (P.A.) of $-50\degr$, which coincides with that found by \citet{smith02} based on symmetry arguments. We suggest that these holes must form a fundamental axis of the Homunculus, which could be created because of a low (or even inhibited) mass-loss rate within $\approx 5\degr$ of the poles. Since about 75 per cent of the mass of the Homunculus is located at high stellar latitudes \citep{smith06}, when the lobes expand they might appear as two rings in the future, similar to those seen around other blue supergiants, such as HD168625 or Sher~25 \citep{smith07}.

An alternative explanation is that the central star has had a major blowout in the polar region, creating the holes. In this scenario, this `blowout' would have been a greater manifestation of the same mechanism that produced the many fast-moving structures dubbed `strings' or `whiskers' or even `spikes' \citep{weisetal99,morseetal98,meaburnetal96}. These high-density ($n_{\rm{e}}\sim10^4$~cm$^{-3}$; \citealt{weis02}) filamentar structures lie outside the Homunculus and are moving at nearly 1000~km\,s$^{-1}$ but even so, they do not emit hard X-rays most likely because of its very small cross-section. Interestingly, they are only seen at high stellar latitudes (in the polar directions). One of the explanations for the observed velocity profile (a Hubble law) of these structures is that they could be formed in a presumably stellar explosion \citep{weisetal99}. Thus, if this scenario is correct, this explosion could be responsible for the formation of the hole in pole as well.

We also note that an explosion at the surface of the primary star was the physical mechanism used by \citet{smithetal07iii} in a simulation that creates, simultaneously, a bipolar nebula and an equatorial disc (as observed in the Homunculus). However, the physical mechanism that could start such stellar explosion remains unclear, and encourages further studies.

\subsubsection{Spatial mapping of the \textit{Little Homunculus}}

Due to the long-slit spectroscopic technique employed by \citet{smith05}, the determination of the spatial extent and distribution of the \textit{Little Homunculus} (hereafter LH) was restricted to the interpolation between the points where the emission associated with the LH was detected in the slit. In the present work, we show the 3D kinematics in the form of slices in velocity space, rather than slices along the major axis as in \citet{smith05}, but the results of the two independent methods are in agreement. Our velocity channel images may provide a better way to evaluate images from simulations of the formation of the LH \citep{gonzalezetal04}, as we provide a complete, model-independent spatial map of the LH.

\begin{figure*}\centering \vbox {\vfil \includegraphics[width=16.8cm]{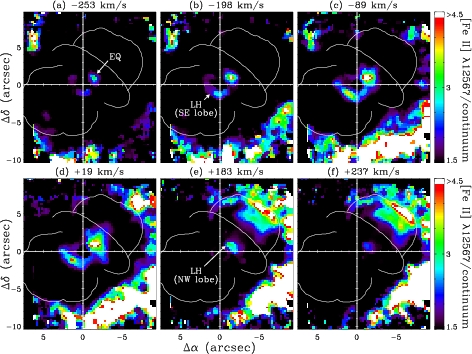} \caption{\label{fig5}Velocity maps of the [Fe\,{\sc ii}] $\lambda$12567 line showing the observed structures and its identification. LH means emission associated to the \textit{Little Homunculus} while EQ is due to the equatorial disc. A fully animated version of this figure is available at \href{http://www.astro.iag.usp.br/~damineli/feii_lh.gif}{www.astro.iag.usp.br/$\sim$damineli/feii\_lh.gif} with the same color scale.}\vfil} \end{figure*}

Our analysis of the velocity maps (Fig.~\ref{fig5}) showed that the emission of the SE lobe of the LH begins at~$\approx -250$~km~s$^{-1}$ and goes up to $+100$~km~s$^{-1}$. The emission associated to the LH is seen blue-shifted near the centre in the SE lobe as indicated in Fig.~\ref{fig5}(a) and (b), in line with the results of \citet{smith05}. Starting from negative and moving toward positive velocities, it is possible to see the emission from the equatorial disc (EQ in Fig.~\ref{fig5}(a)) in the same line-of-sight of the NW lobe of the LH. However, based on geometric arguments, the components of the equatorial disc can only have negative velocities \citep{davidsonetal97b}. Thus, the component associated to the NW lobe of the LH was identified as the structure lying near the central region with velocities ranging from $\approx+20$ up to about $+235$~km\,~s$^{-1}$ (see Fig.~\ref{fig5}(d)--(f)).

\subsection{The H\lowercase{e\,{\sc i}} $\lambda$10830 emission column}\label{beam}

The He\,{\sc i}~$\lambda$10830 line has a very complex velocity structure in the spectrum of the Homunculus \citep{smith02}. It is a combination of absorption, emission and reflection from different regions inside and outside the nebula and is, presumably, formed near the central source.

\begin{figure}\centering \includegraphics[width=8.4cm]{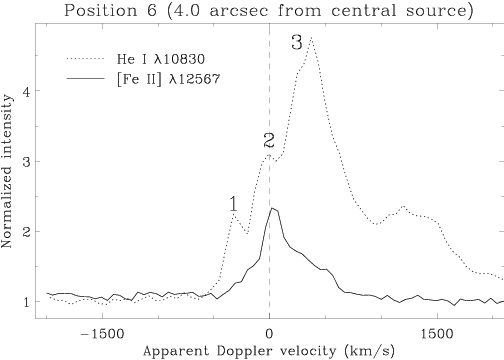} \caption{\label{fig6}Comparison between He\,{\sc i}~$\lambda$10830 and [Fe\,{\sc ii}]~$\lambda$12567. The components labelled 1 and 2 are intrinsic emission from the equatorial region while 3 is formed in the winds of the central source and then scattered by the dust in the background NW lobe. Both spectra were extracted from the point labelled 6 in Fig.~\ref{fig7}. The broad line at $\approx1200$ ~km~s$^{-1}$ is a blend between Fe\,{\sc ii} $\lambda$10863 and Fe\,{\sc ii} $\lambda$10872.} \end{figure}

We detected an intrinsic emission component which appears restricted to a narrow azimuthal region in the line-of-sight to the NW lobe \citep{smith02}. It is labelled as 1 in Fig.~\ref{fig6}, and is likely photo-excited by energetic photons (at least $16.2$~eV), since [Fe\,{\sc ii}]~$\lambda$12567 does not show any component at the same velocity. Component 2 is associated with emission from the slow-moving equatorial ejecta\footnote{Note, however, that component 2 may also be associated with diffuse emission due to the H\,\textsc{ii} region in which $\eta$~Carinae is immersed.}, and component 3 is reflected emission from the central source in the NW lobe (see also fig.~12 and 13 of \citealt{smith02}).

\begin{figure}\centering \vbox {\vfil \includegraphics[width=8.4cm]{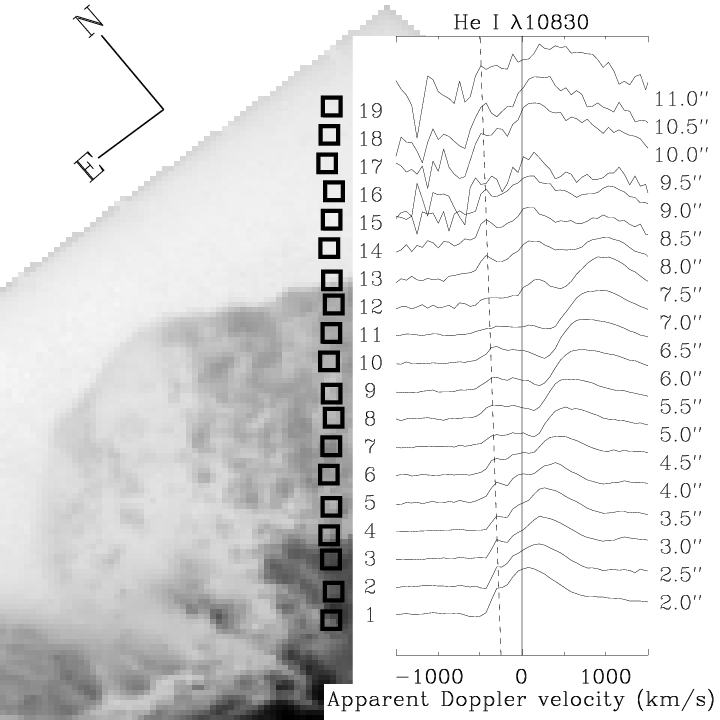} \caption{\label{fig7}He\,{\sc i}~$\lambda$10830 line profile along position angle $-41\degr$. Notice the detection of the He\textsc{i}~$\lambda$10830 emission column along this position angle (vertical dashed line). In order to improve the signal-to-noise ratio, each spectrum was median-combined from 4 adjacent lens. The backgound image of the Homunculus (from \citealt{morseetal98}) was rotated by $41\degr$ counter-clockwise.
}\vfil} \end{figure}

This narrow intrinsic emission of He\,{\sc i}~$\lambda$10830 (hereafter He emission column) is likely formed when the UV radiation from the central source passes through the holes in the torus \citep{smith02,smithetal02} and are free to excite He atoms at large radii from the central source. The apparent Doppler velocity of the He emission column changes, respectively, from $\approx-250$ to $\approx-500$~km~s$^{-1}$ when moving from a projected distance of 2~arcsec to 10.5~arcsec from the central source (see Fig.~\ref{fig7}), suggesting that the emitting region has a Hubble-flow motion, i.e. $v\propto d$. Note that such high projected velocity would suggest that the equatorial disc has ejecta moving as fast as the polar cap of the Homunculus. However, as will be discussed in the next section, given the errors associated to the determination of the inclination angle of the helium emission column (which indicates whether or not it is in the equatorial disc), it may be comfortably associated with the Great Eruption, which lasted about 20 yr with peak between 1843 and 1851 \citep{currieetal96,currieetal99,morseetal01}.

\subsubsection{Characteristics of the helium emission column and the detection of its twin brother}

\begin{figure*}\centering \vbox {\vfil \includegraphics[width=16.8cm]{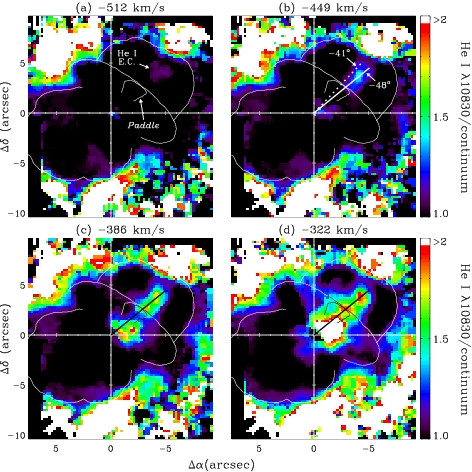} \caption{\label{fig8}Velocity maps for the He\,{\sc i}~$\lambda$10830 line showing the location of the He emission column (He~\textsc{i}~E.C. in the figure) and the \textit{Paddle}. The misalignment between the P.A. of the \textit{Paddle} and that of the helium emission column is clearly seen in (b). Also, note the strong emission (white region) near the centre in (d), which is associated to the \textit{Purple Haze}. A full animated version of this figure is available at \href{http://www.astro.iag.usp.br/~damineli/hei.gif}{www.astro.iag.usp.br/$\sim$damineli/hei.gif} with the same color scale.}\vfil} \end{figure*}

It is known that the \textit{Paddle} is a dust-free region located at the equatorial disc \citep{smithetal98}. Its shape is well defined and seems symmetric regarding the major axis of the Homunculus\footnote{Although it is likely that this alignment occurs by chance.} (see Fig.~\ref{fig1}). Therefore, the \textit{Paddle} would be a suitable candidate to be blamed for the escape of radiation. Indeed, free of any interaction, it is expected that radiation flowing through that region would follow a linear path centered on P.A.=$-41\degr$ -- the position angle of the \textit{Paddle}. However, the observed P.A. of the He\,\textsc{i} emission column is different of the \textit{Paddle}, namely the He emission is at $-48\degr$ (see Fig.~\ref{fig8}). We must stress that even with this discrepancy, the He\textsc{i}$\lambda$10830 emission column can be weakly detected along P.A.=$-41\degr$ because of its roughly 3~arcsec wide but the bulk of emission comes indeed from P.A.=$-48\degr$.

In order to analyse the kinematic structure of the He\,\textsc{i}$\lambda$10830, we adopted the same convention as in \citet{davidsonetal01} and calculated the inclination angle\footnote{The inclination angle is defined such that $i=0\degr$ means that the equatorial disc is seen face-on, while $i=90\degr$ corresponds to an edge-on view.} of the He emission column using the following equation
\[ \tan(i_{He})=\frac{t}{4.74 D}\frac{V}{\omega}, \]
where $t$ is the age, in years, of the equatorial gas, $D$ is the heliocentric distance to $\eta$~Car measured in pc, $V$ is the apparent Doppler velocity (km~s$^{-1}$) measured at position $\omega$ (arcsec). Note that regarding the age of the equatorial disc, there is no consensus. Based on kinematics studies, \citet{morseetal01} suggested that the equatorial disc is coeval with the Homunculus lobes, although some material appears to be even younger presumably associated with posterior eruptions \citep{davidsonetal97,davidsonetal01,smithetal98a,dorlandetal04}. In the present work, we assumed an age of 160 year for the equatorial disc. We also assumed that it is perpendicular to the major axis of the Homunculus \citep{davidsonetal01,smithetal07iii}.

From our data, the average value of $|V/\omega|$ was $67.9\pm6.3$~km\,s$^{-1}$~arcsec$^{-1}$, corresponding to $\tan(i_{He})=0.95\pm0.13$, which in turn results an inclination angle of approximately $44\degr$ \textit{from the plane of the sky toward us}. Though the error in our result ($\pm3\degr$) is larger than that obtained with long-slit observations -- typically less than 1\degr --, the lower end is consistent with $i=41\degr$, which is the assumed inclination angle of the equatorial disc obtained using long-slit observations. Even so, our range of values for the inclination angle of the helium emission column is coherent with the Great Eruption, which takes place between 1837 and 1860, with the peak occuring around 1843 and 1851 \citep{currieetal96,currieetal99,morseetal01}.

Hence, we concluded that, if the He\,\textsc{i}$\lambda$10830 emission column is indeed in the equatorial disc and is caused by UV escaping through a hole in the torus, we should detect this same effect in other places, since it is known that there are many holes in the equatorial torus.

\begin{figure*}\centering \vbox {\vfil \includegraphics[width=16.8cm]{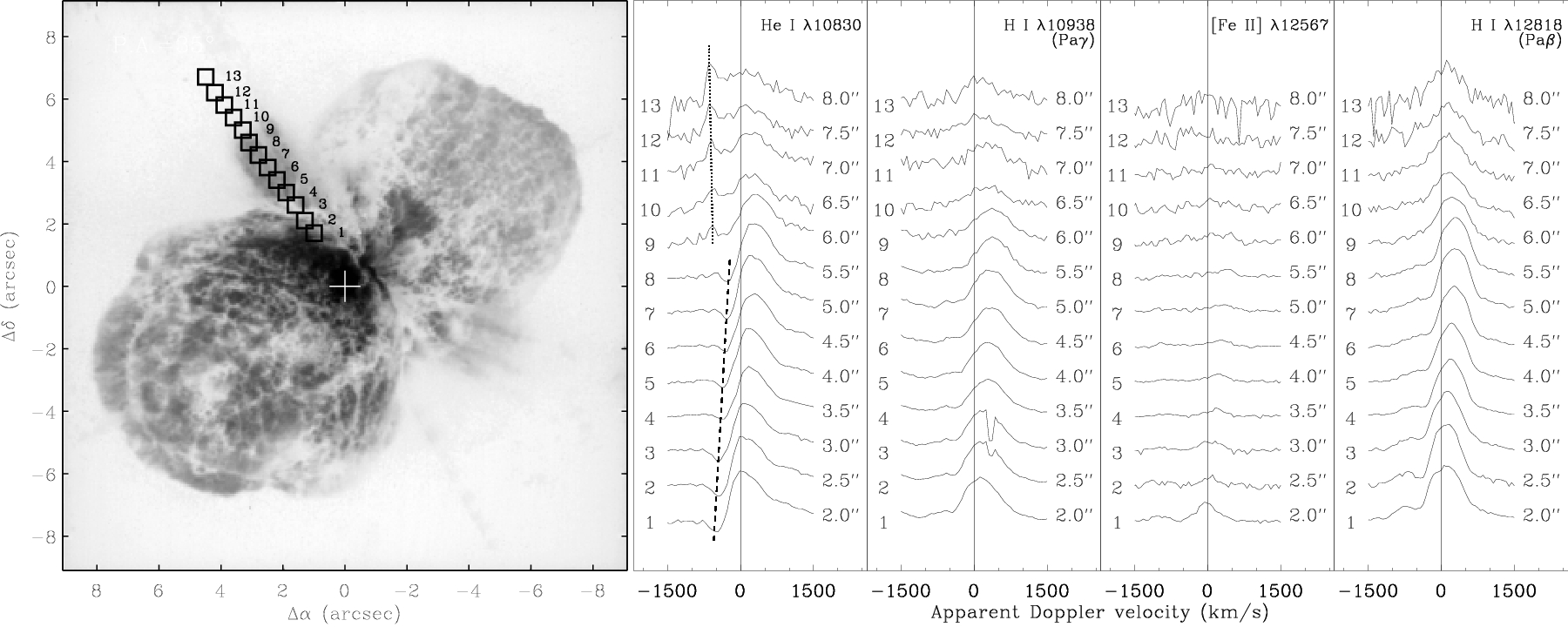} \caption{\label{fig9}Tracing of line profiles toward the \textit{NN jet}. Notice the changes in the P Cygni absorption component in the helium line (dashed line), which moves from approximately $-500$ to $-220$~km\,s$^{-1}$ from position 1 to 9, respectively. Beyond position 9, we detected another He\textsc{i}~$\lambda$10830 emission column (dotted line). At position 2 and 3, Pa$\gamma$ shows an artifact in the red side of the line that is caused by a cosmetic defect in the detector. The background image is from \citet{morseetal98}.}\vfil} \end{figure*}

In fact, we also detected intrinsic emission of He\textsc{i}$\lambda$10830 at the end of the \textit{NN jet} (Fig~\ref{fig9}). With a P.A. of +35\degr, the spectrum extracted from 2 up to 6~arcsec from the central source is likely reflected by dust in the \textit{NN jet}, since they show a red-shifted velocity profile where it is expected to find only blue-shifted velocities (if they were due to intrinsic emission). Furthermore, He\,\textsc{i}$\lambda$10830 shows a velocity-variable P~Cyg profile in that region (shown by the dashed line from position 1 to 8 in Fig.~\ref{fig9}), which is also a strong suggestion of reflection. However, beyond 6~arcsec from the central source, the reflected profile disappears and a blue-shifted emission begins to raise at approximately $-570$~km\,s$^{-1}$ going up to $-650$~km\,s$^{-1}$ at 8~arcsec (see Fig.~\ref{fig9}). Therefore, our results suggest that there are at least two regions where radiation is escaping to excite/ionize gas lying in the equatorial disc.

In a binary context, these regions could be produced by the high-energy radiation coming direct from the hot source of the system, which spends most of its orbital period near apastron, in a highly elliptical orbit. Thus, if the plane of the orbit is the same as the equatorial disc, then one would expect the UV from the secondary to escape through the holes in the torus and excite/ionize helium atoms along its way. It would be very interesting to observe these regions along the period of 5.52 year to see their behavior near the minimum, when the hot companion gets into the dense wind of the primary. Hence, if our assumption is correct, the He\,\textsc{i}$\lambda$10830 emission column would fade and then return.

\begin{figure*}\centering \vbox {\vfil \includegraphics[width=16.8cm]{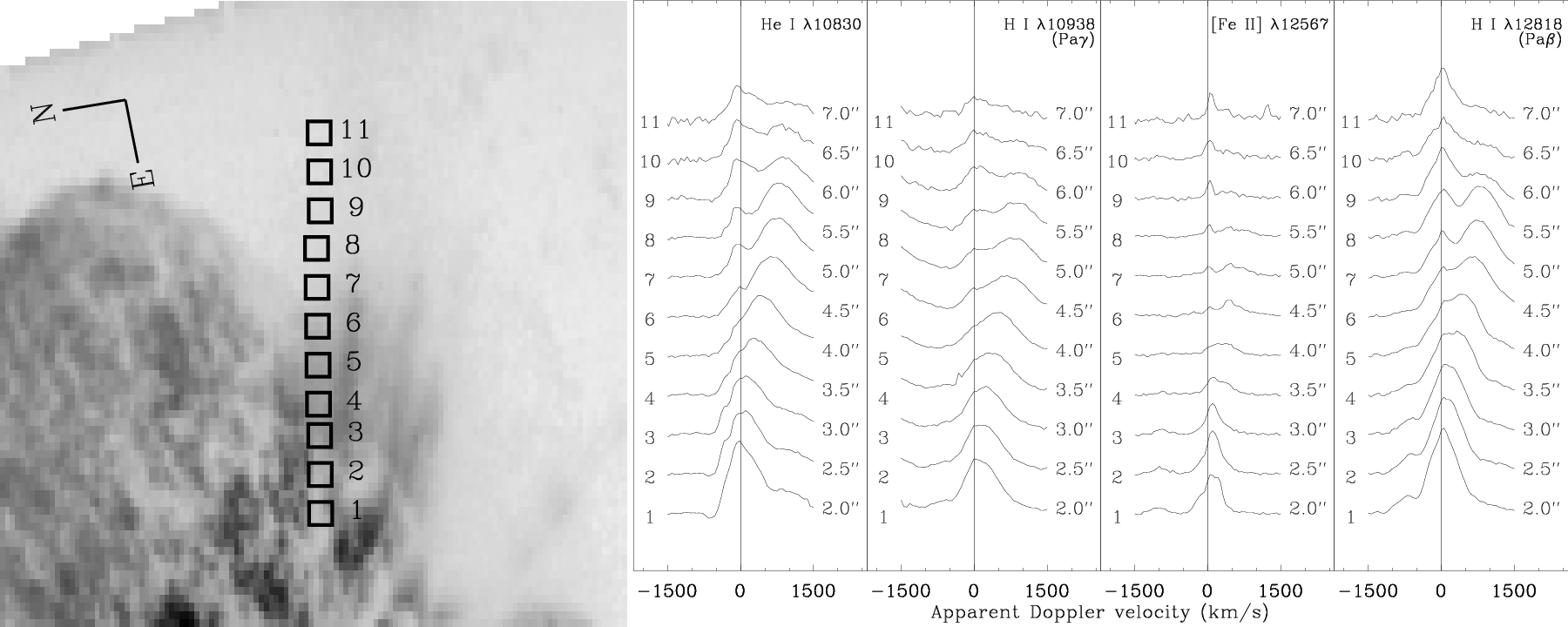} \caption{\label{fig10}Same as Fig. \ref{fig9} but extracted towards the \textit{S-condensation}. At this position angle, we did not detected any component in the helium line that could be due to high-energy photons (far-UV from the secondary) escaping through holes in the equatorial torus. We only detected intrinsic emission due to the equatorial disc as well as reflected components. Notice that all of the lines show a component near 0~km\,s$^{-1}$, which remains with practically the same apparent Doppler velocity at all of the positions. This component may be associated to the H\textsc{ii} region in which $\eta$~Car is immersed. To better visualization, the image of the Homunculus was rotated by $-100\degr$.}\vfil} \end{figure*}

We also noted that to the SW direction -- toward the \textit{S-condensation} --, we only detected reflected components as well as intrisic emission associated with the equatorial disc but no signal of another He\,\textsc{i}$\lambda$10830 emission column (see Fig~\ref{fig10}).

\section{Discussion}\label{dis}

\subsection{The 3-cm radio emission}\label{radio}

\begin{figure}\centering \includegraphics[width=8.4cm]{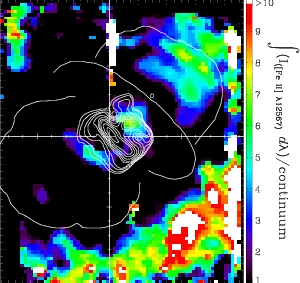} \caption{\label{fig11}The background image represents the integral of the [Fe\,{\sc ii}]~$\lambda$12567 line flux from $-1000$ to $+1000$~km\,s$^{-1}$ normalized by the adjacent continuum. The continuum radio-emission at 3~cm (from \citealt{duncanetal97}) is superimposed to show that the low levels match the spatial extent of the \textit{Little Homunculus}, as claimed by \citet{smith05}. Also, note the spatial coincidence between the NW radio peak and and the emission from the equatorial region (see discussion in \S\ref{radio}).} \end{figure}

\begin{figure}\centering \vbox {\vfil \includegraphics[width=8.4cm]{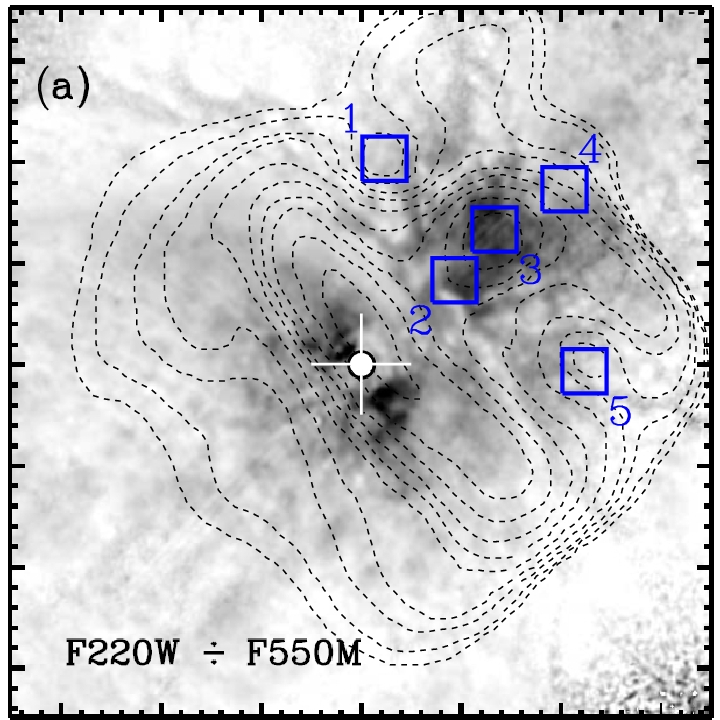} \includegraphics[width=8.4cm]{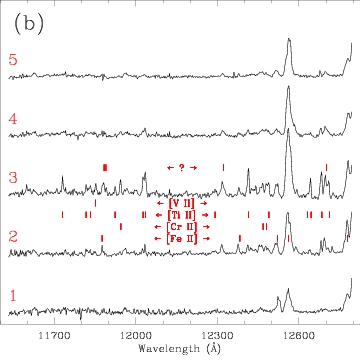} \caption{\label{fig12}(a) 3-cm radio contours from \citet{duncanetal97} superimposed on an image of the ratio of F220W to F550M from \citet{smithetal04} (reproduced with permission). In the background image, dark areas indicate stronger UV emission. The radio contours were scaled to take into account an expansion due to the 7-year delay between the radio and optical observations. (b) The spectrum taken from the regions indicated in (a) shows the presence of several low-ionization lines at positions 2 and 3, which have stronger intensity as compared to 1, 4 and 5.}\vfil} \end{figure}

\begin{table}\centering
\caption{\label{t1}Emission lines observed in the region of the `Sr-filament'. The unidentified lines are listed with a `?' symbol. Note that [Fe\,\textsc{ii}]~$\lambda$12567 has three components (shown in bold font).}
\label{lines}
\begin{tabular}{@{}ccc}
\hline
$\lambda_{\rm{obs}}$ (\AA) & Ion & Velocity (km\,s$^{-1}$) \\
\hline
11730.8 & [Ti\,{\sc ii}]~$\lambda$11736 & $-130$ \\
11817.7 & [Ti\,{\sc ii}]~$\lambda$11823 & $-142$ \\
11833.8 & [Ti\,{\sc ii}]~$\lambda$11838 & $-132$ \\
11853.0 & [V\,{\sc ii}]~$\lambda$11857  & $-110$ \\
11877.2 & [Fe\,{\sc ii}]~$\lambda$11882  & $-110$ \\
11883.9 & ?                             &        \\
11890.8 & ?                             &        \\
11925.2 & [Ti\,{\sc ii}]~$\lambda$11930 & $-130$ \\
11945.8 & [Cr\,{\sc ii}]~$\lambda$11950 & $-115$ \\
12028.1 & [Ti\,{\sc ii}]~$\lambda$12033 & $-115$ \\
12037.1 & [Ti\,{\sc ii}]~$\lambda$12042 & $-115$ \\
12292.8 & [Ti\,{\sc ii}]~$\lambda$12298 & $-115$ \\
12322.9 & ?			        &        \\
12383.5 & [Fe\,{\sc ii}]~$\lambda$12388 & $-100$ \\
12416.6 & [Ti\,{\sc ii}]~$\lambda$12422 & $-120$ \\
12469.7 & [Cr\,{\sc ii}]~$\lambda$12476 & $-140$ \\
12482.8 & [Cr\,{\sc ii}]~$\lambda$12488 & $-135$ \\
12490.5 & [Ti\,{\sc ii}]~$\lambda$12496 & $-140$ \\
12521.8 & [Fe\,{\sc ii}]~$\lambda$12521 & $+10$  \\

\bf{12556.3} & \multirow{3}{*}{\bf{[Fe\,{\sc ii}]~$\lambda$12567}} & \bf{$-$250} \\
\bf{12563.2} &    & \bf{$-$85} \\
\bf{12573.1} &    & +\bf{150} \\

12633.0 & [Ti\,{\sc ii}]~$\lambda$12638 & $-120$ \\
12646.5 & [Ti\,{\sc ii}]~$\lambda$12651 & $-105$ \\
12685.5 & [Ti\,{\sc ii}]~$\lambda$12692 & $-145$ \\
12703.1 & ?                             &        \\
12713.7 & [Ti\,{\sc ii}]/[Mn\,{\sc ii}]~$\lambda$12719 & $-115$ \\
12787.3 & [Fe\,{\sc ii}]~$\lambda$12787 & $+10$  \\ \hline
\multicolumn{2}{c}{mean velocity} &  $-109\pm9$\\
\hline
\end{tabular}
\end{table}

The radio monitoring performed by \citet{duncanetal03} revealed that during the low-excitation phases (when the high-excitation lines weaken or vanish -- \citealt{gaviola53,rodgersetal67,thackeray67,zanellaetal84}) the free-free emission is concentrated in a small region of $\sim 1.5 $~arcsec in diameter. However, during the `normal' state, a more extended 3-cm radio emission region of about 4~arcsec is present, which can be seen by the contours in Fig.~\ref{fig11}.

It is often assumed that the structure seen at 3-cm radio continuum is the equatorial torus. In this context, during the high-excitation state, the surrounding torus absorbs the UV radiation and then is kept ionized throughout the most part of the orbital period. However, when the secondary star is at periastron, the ionizing flux is rapidly absorbed by the dense wind of the primary star and thus the previously-ionized 4~arcsec-wide region -- namely, the torus -- is allowed to recombine and therefore, the radio-continuum flux is reduced to a point-like source, i.e. restricted to a small Str\"omgren sphere around the hot companion.

On the other hand, the ionizing flux could \textit{also} be absorbed in the lobes of the LH. \citet{duncanetal97} showed that the H91$\alpha$ flux is composed of mainly two components: one bright and narrow feature (FWHM$\approx250$~km\,s$^{-1}$) with peak at $-250$~km\,s$^{-1}$ and a broad (FWHM$\approx600$~km\,s$^{-1}$), fainter emission with peak at approximately $-115$~km\,s$^{-1}$ \citep{duncanetal97}. The brightest component is presumably equatorial emission due to a turbulent gas cloud located at 1.6~arcsec NW of the central source. This cloud must be colder than the rest of the torus because of its high ratio of spectral-line to continuum, which is an indicator of the temperature of the emitting region (\citealt{duncanetal97} and references therein).

Although the projected position of this cold cloud (which we refer to as the radio spot) is the same as the `Sr-filament' (see Fig.~\ref{fig12}), they obviously do not occupy the same spatial region, since the radio spot must be ionized to be seen in radio frequencies while the `Sr-filament' shows many lines of low-ionization ions such as [Sr\textsc{ii}], [Ti\textsc{ii}], etc. (\citealt{zethsonetal99,hartmanetal01,hartmanetal04}). Lines of these same ions were also present in our data, as shown in Fig.~\ref{fig12} and listed in Table~\ref{t1}. In addition, the kinematics of the `Sr-filament' ($\approx-100$~km\,s$^{-1}$) \textit{does not match} the observed velocity of the peak of the bright H91$\alpha$ component ($\approx-250$~km\,s$^{-1}$). We noted, though, that the He\,\textsc{i}$\lambda$10830 emission column \textit{does match} the velocity of the radio peak. Moreover, they are located at roughly the same projected position, suggesting that they may be related to each other. There is no doubt that this relationship deserves further studies, since it could give important clues about the ionizing source.

Regarding the faint component of the H91$\alpha$ emission, it could arise from both the torus and the LH because of the spatial coincidence that exists between the low-level contours of the broad radio-emission and the extent of the LH (see Fig.~\ref{fig11}). Moreover, the FWHM of this component is similar to the range of velocities observed in the LH ($\approx\pm250$~km\,s$^{-1}$; \citealt{smith05}).

\subsection{On the nature of $\bmath{\eta}$ C\lowercase{ar} B}\label{secondary}

Independently of where the radio emission is formed and assuming that the free-free emission discussed in \S~\ref{radio} is caused mainly by the UV radiation field of $\eta$~Car~B, we estimated the number of Lyman continuum photons using the following relation \citep{mezgeretal74,carpenteretal90,filipovicetal03,morganetal04} \[ N_{\rmn{Ly}}=\phi\hspace{1mm}a(\nu,T_{\rmn{e}})^{-1} \left[\frac{\nu}{\rmn{GHz}}\right]^{0.1}\left[\frac{T_\rmn{e}}{\rmn{K}}\right]^{-0.45} \left[\frac{S_\nu}{\rmn{Jy}}\right]\left[\frac{D}{\rmn{kpc}}\right]^2 \] where $N_{\rmn{Ly}}$ is the number of photons per second in the Lyman continuum, $\phi$ is a numeric constant (=~4.76~$\times~10^{48}$), $a(\nu,T_{\rmn{e}})$ is a slowly varying function tabulated by \citet{mezgeretal67}, $\nu$ is the frequency at which the observation is made, $T_{\rmn{e}}$ is the electronic temperature, $S_\nu$ is the flux density, and $D$ is the distance to the source.

Using the observed radio continuum flux at 1.3~mm\footnote{At this wavelength the free-free emission is optically thin \citep{abrahametal05}.} when the system is at high-excitation state (i.e., when $\eta$~Car~B is out of dense wind of the primary star) $S_\nu\approx~39$~Jy (Abraham, Z., private communication) and a value of $a(\nu,T_\rmn{e})\sim1$ for $T_\rmn{e}\sim10^{4}$~K \citep{mezgeretal67}, the ionising flux is log($N_\rmn{Ly})\approx49.4$. This corresponds to a star with the minimum spectral type  of  O5.5\,III to O7\,I star \citep{martinsetal05}. As we can only put a lower limit to the number of Lyman continuum photons and to the spectral type, we can not rule out a late-type Wolf-Rayet companion, which also has log($N_\rmn{Ly})>49.4$ \citep{crowther07}.

An O7\,I presents stellar parameters compatible with other other works. The effective temperature for such a star is about 35,000~K \citep{martinsetal05}, which is well within the range of 34,000--38,000~K determined by \citet{verneretal05} based on the observed ratio of [Ar\,{\sc iii}]~$\lambda7136$ to [Ne\,{\sc iii}]~$\lambda3869$ in the Weigelt blobs. \citet{ipingetal05} also indicates an effective temperature near 35,000~K using spectra obtained with the \textit{Far Ultraviolet Spectroscopic Explorer} (\textit{FUSE}). Altough the flux of $\eta$~Car~B is expected to dominate the spectra at wavelengths shortward of 1200~\AA~\citep{hillieretal06}, the quantitative analysis is rather complex \citep{hillieretal06}. An O5.5\,III star presents stellar parameters that are also similar to those assumed for $\eta$~Car~B. Following \citet{martinsetal05}, an O5.5\,III star has an effective temperature of about 39,250~K, which is higher than the upper limit of \citet{verneretal05}. Nevertheless, this effective temperature is rather acceptable since, there may be gas between the ionizing source and the Weigelt blobs responsible for absorbing the high energy part of the spectrum, and thus decreasing the estimating of the effective temperature.

It is worthwhile to note that \citet{prinjaetal90} showed that early-type stars in the range from O5.5 ($T_{\rmn{eff}}\approx~39,250$~K) to O7 ($T_{\rmn{eff}}\approx35,000$~K) present terminal velocities in the range from 1100 up to 3000~km~s$^{-1}$, which agrees with the values proposed for $\eta$~Car~B by \citep{davidson99,pittardetal03} based on hydrodynamic calculations.

\section{Summary and conclusions}\label{sum}

Near-infrared integral field spectroscopy has revealed additional details of the circumstellar ejecta around $\eta$ Car. The main results and conclusions are summarized below:

\begin{enumerate}

\item We determined the dimensions and geometry of the hole present in both lobes of the Homunculus using our [Fe\,{\sc ii}]~$\lambda$12567 velocity maps and the model of \citet{smith06} for the Homunculus. The holes have a diameter of $\approx6.0\times10^{16}$~cm and is $\approx6.5\times10^{16}$~cm-thick at the polar region. They are located within $5\degr$ from the pole, suggesting an inhibited mass loss at stellar latitudes $\ga 85\degr$ during the Great Eruption. These holes are seen in the H$_2$ and [Fe\,{\sc ii}]~$\lambda$16435 lines as well \citep{smith06};

\item The feature known as the \textit{SE hole} in optical images is a region with a local minimum column density towards the SE lobe caused by the fact that we are looking through the borders of the hole;

\item We confirmed the claim of \citet{smith05} and also suggested that the broad component of the 3-cm continuum radio-emission originates both in the torus and the LH because of the spatial coincidence between the low-level contours of the radio emission and the extent of the LH, though the bulk of emission is due to the torus. Moreover, the width of the broad radio-emission is also consistent with the kinematics of the LH.

\item The He\textsc{i}$\lambda$10830 emission column presents a Hubble flow from $-250$~km\,s$^{-1}$ (at 2~arcsec from the central source) to $\approx-500$~km\,s$^{-1}$ (at 10.5~arcsec). Its position angle is $-48\degr$ and based on symmetry and kinematic arguments, we suggested that the He\textsc{i}$\lambda$10830 emission column is not related to the \textit{Paddle}, which shows P.A.=$-41\degr$. Nonetheless, our results suggest that it is indeed in the equatorial disc (with inclination angle of $i_{He}=44\degr$) and is most likely related to the radio spot (the narrow component of the 3-cm continuum radio-emission reported by \citealt{duncanetal97}).

\item We also detected another He\,\textsc{i}~$\lambda$10830 emission column at P.A.=$+35\degr$, confirming the suggestion that such structure is indeed caused by high-energy photons (far-UV from $\eta$~Car~B) escaping through holes in the equatorial disc.

\item The radio spot and the `Sr-filament' are in the same line-of-sight but disconnected spatially. While the former is an ionized region with peak at $-250$~km\,s$^{-1}$, the latter is characterized by low-ionization lines with typically $-110$~km\,s$^{-1}$, presumably shielded from high-energy radiation by H$^{0}$ and a forest of Fe$^{+}$ \citep{bautistaetal02,bautistaetal06}.

\item From the observed 1.3-mm radio flux we estimated that the ionising flux, which comes from $\eta$~Car~B, is consistent with an O-type star hotter than O5.5\,III to O7\,I, though we can not rule out a Wolf-Rayet nature to the companion at this point.

\end{enumerate}

\section{Acknowledgments}

M. Teodoro, A. Damineli, J. H. Groh and C. L. Barbosa are grateful to the Brazilian agencies CNPq and FAPESP for continuous financial support. We would like to thank Dr. Nathan Smith for his extensive comments and suggestions on the earlier stages of this manuscript. We are grateful to the referee Dr. Theodore Gull for his fruitful comments that have improved the content and presentation of our results. M. Teodoro also would like to thank Michelle Doherty for her efforts in obtaining all the calibration data as well as Dr. Jon Morse for kindly granting the permission to use the \textit{HST}/WFPC2 images of the Homunculus shown in this paper. M. Teodoro is supported by FAPESP through grant 05/00190-8.

\bibliography{refs}

\bsp

\label{lastpage}

\end{document}